\documentclass{jfm}
\usepackage{graphicx}
\usepackage{natbib}
\usepackage{hyperref}
\usepackage{newtxtext}
\usepackage{newtxmath}
\usepackage{tabularx}
\usepackage{amsmath}
\usepackage{bm}
\usepackage[geometry]{ifsym}
\usepackage{subcaption}
\usepackage{comment}
\usepackage[normalem]{ulem}
\usepackage[T1]{fontenc}

\usepackage{xcolor}

\colorlet{houssem}{black}%purple}
\colorlet{shuai}{black}
\colorlet{anubhab}{black}

\newcommand{\Sto}{\mathrm{St}_\Gamma}

\title{The merger of co-rotating vortices in dusty flows}

\shortauthor{S. Shuai, A. Roy, and M. H. Kasbaoui}

\author{
  Shuai Shuai \aff{1} and
  Anubhab Roy\aff{2} and
  M. Houssem Kasbaoui \aff{1} \corresp{\email{houssem.kasbaoui@asu.edu}}
}

\affiliation{
  \aff{1}School for Engineering of Matter, Transport and Energy, Arizona State University, Tempe, AZ 85281, USA.
  \aff{2}Department of Applied Mechanics, Indian Institute of Technology Madras, Chennai, 600036, India.
}
\begin{document}

\maketitle

\begin{abstract}
We investigate the effect of particle inertia on the merger of co-rotating dusty vortex pairs at semi-dilute concentrations. {\color{houssem}In a particle-free flow, the merger is triggered once the ratio of vortex core size to vortex separation reaches a critical value. The vortex pair separation then decreases monotonically until the two cores merge together. Using Eulerian-Lagrangian simulations of co-rotating particle-laden vortices, we show substantial departure from the vortex dynamics previously established in particle-free flows.} 
Most strikingly, we find that disperse particles with moderate inertia cause the vortex pair to push apart to a separation nearly twice as large as the initial separation. \textcolor{houssem}{During this stage, the drag force exerted by particles ejected out of the vortex cores on the fluid results in a net repulsive force that pushes the two cores apart.}
Eventually, the two dusty vortices merge into a single vortex with most particles accumulating outside the core similar to the dusty Lamb-Oseen vortex described in Shuai \& Kasbaoui (\emph{J. Fluid Mech.}, vol 936, 2022, A8)
For weakly inertial particles, we find that the merger dynamics follow the same mechanics as those of a single-phase flow, albeit with a density that must be adjusted to match the mixture density. 
\textcolor{houssem}{For highly inertial particles, the feedback force exerted by the particles on the fluid may stretch the two cores during the merger to a point where each core splits into two, resulting in inner and outer vortex pairs. In this case, the merger occurs in two stages where the inner vortices merge first, followed by the outer ones.}
\end{abstract}


\section{Introduction}
{\color{houssem}
The merger of vortices is relevant to many engineering problems and practical applications including aeronautics, geophysical fluid dynamics, meteorology, and astrophysics \citep{robertsTopicsComputationalFluid1972,rossowConvectiveMergingVortex1977,overmanEvolutionMergerIsolated1982}. Much of the prior work on co-rotating vortices was motivated by the observation of vortical structures in aircraft trailing wakes \citep{jacquinUnsteadinessInstabilityTurbulence2005,chenDynamicsCorotatingVortex1999}, which represent a significant hazard to following aircrafts during take-off and landing. Vortex merger is also relevant in many geophysical flows where the large scale motions appear as two-dimensional turbulence \citep{boffettaTwoDimensionalTurbulence2012}.  In such flows, vortex merger is the primary mechanism for the evolution of the flow, as the merger of small scale vortices produces larger vortices and leads to the transfer of energy to larger scales, a well-known characteristic of two-dimensional turbulence \citep{jimenezStructureVorticesFreely1996,couderHydrodynamicsSoapFilms1989,mcwilliamsVORTICESTWODIMENSIONALTURBULENCE1990,hopfingerVorticesRotatingFluids2003}. In three-dimensional turbulence, vortex interaction occurs between coherent flow structures in the region of like-signed vorticity \citep{vincentSpatialStructureStatistical1991,cadotCharacterizationLowPressure1995}.

The majority of prior work on vortex merger concerned single-phase flows. In particular, the two-dimensional dynamics of an identical co-rotating vortex pair in a particle-free flow have been studied extensively \citep{griffithsCoalescingGeostrophicVortices1987,melanderSymmetricVortexMerger1988,waughEfficiencySymmetricVortex1992,dritschelGeneralTheoryTwodimensional1995,meunierMergingCriterionTwodimensional2002a,cerretelliPhysicalMechanismVortex2003a,brandt2006physics,orlandi2007two}. The current understanding is that two co-rotating vortices with equal strength do not merge until the ratio of vortex core size $a$ to pair separation $b$ exceeds a critical value. If the $a/b$ is below the critical threshold $(a/b)_\mathrm{crit}$, the two vortices undergo a diffusive stage during which their sizes grow by viscous diffusion as they rotate around one another, but their separation remains approximately constant. This is generally referred to as the \emph{first diffusive} stage \citep{cerretelliPhysicalMechanismVortex2003a,meunierMergingCriterionTwodimensional2002a}. The \emph{convective stage} starts once $(a/b)$ reaches the critical threshold $(a/b)_\mathrm{crit}$ during which the vortex separation decreases significantly. In one of the earliest merger experiments, \citet{griffithsCoalescingGeostrophicVortices1987} found  $(a/b)_\mathrm{crit}$ to be about $\sim 0.29-0.32$. Later experiments by \citet{meunierThreedimensionalInstabilityVortex2001a} and \citet{cerretelliPhysicalMechanismVortex2003a} showed that the threshold is closer to $(a/b)_\mathrm{crit}= 0.29$.  For vortex pairs with nonuniform vorticity distributions, \citet{meunierMergingCriterionTwodimensional2002a} proposed a similar merger criterion based on a refined definition of vortex core size. In addition to experimental observation, linear stability analyses carried out by \citet{dritschelGeneralTheoryTwodimensional1995} and \citet{meunierMergingCriterionTwodimensional2002a} show that vortex pairs that are too close, with $a/b>0.32$ in \citep{dritschelGeneralTheoryTwodimensional1995}, become unstable with respect to infinitesimal two-dimensional perturbations. \citet{melanderSymmetricVortexMerger1988} proposed that the formation of vortex filaments during the convective stage drives the merger of the vortex pair. However, this notion was later disputed by \citet{brandt2006physics} and \citet{orlandi2007two}. \citet{cerretelliPhysicalMechanismVortex2003a} showed that the dominant physical mechanism during the convective stage is controlled by the antisymmetric vorticity field. The latter induces a velocity that pulls the two cores together resulting in the two vortices becoming intertwined. They further showed that a \emph{second diffusive} stage follows the convective stage. During this final stage, viscous diffusion dominates once more and smoothens the large vorticity gradients resulting from the merger.

}

{\color{houssem}
Particle-free co-rotating vortices may also be subject to three-dimensional instabilities.  \citet{jimenezStabilityPairCo1975} showed that co-rotating vortices are stable to the Crow instability, a long-wavelength instability that is known for destabilizing counter-rotating vortex tubes \citep{crowStabilityTheoryPair1970a}. However, co-rotating vortex tubes may be susceptible to a short-wavelength instability called the elliptic instability \citep{tsaiStabilityShortWaves1976}.
\citet{meunierThreedimensionalInstabilityVortex2001a} showed that the elliptic instability emerges at circulation Reynolds number $\Rey_\Gamma=\rho_f\Gamma/\mu_f$  over $\sim 2000$. \citet{orlandi2007two} performed direct numerical simulations at $\Rey_\Gamma=3000$, mirroring the experiments of \citet{meunierThreedimensionalInstabilityVortex2001a}, and showed that, depending on the initial axial disturbance, the merger dynamics in 3D can be significantly more complex than in 2D. Below, the critical threshold $\Rey_{\Gamma,\mathrm{crit}}\sim 2000$ a pair of co-rotating vortex tubes evolve in a two-dimensional way.
}

To the best of our knowledge, vortex merger in semi-dilute dusty flows has not been investigated. Yet, the dynamics in these flows may deviate considerably from those in particle-free flows. 
{\color{houssem} This is especially true for dusty flows in the semi-dilute regime, where the average particle volume fraction $\phi_{p,0}$ is in the range $O(10^{-6})$ to $O(10^{-3})$. Due to large  solid-to-gas density ratio ($\rho_p/\rho_f=O(10^3)$), the mass loading $M=\rho_p\phi_{p,0}/\rho_f$ is $O(1)$ in semi-dilute dusty flows which leads to significant feedback force from the particles on the fluid, i.e., two-way coupling. In this regime, the disperse phase may cause large flow modulation. Several experiments and simulations of turbulent flows laden with semi-dilute inertial particles attest to this effect \citep{ahmedMechanismsModifyingStructure2000,hwangHomogeneousIsotropicTurbulence2006,hwangTurbulenceAttenuationSmall2006,meyerModellingTurbulenceModulation2012,richterTurbulenceModificationInertial2015,kasbaouiClusteringEulerEuler2019,pengDirectNumericalInvestigation2019,costaNearwallTurbulenceModulation2021,brandtParticleLadenTurbulenceProgress2022,daveMechanismsDragReduction2023}. However, there is a dearth of studies of particle-vortex interaction in canonical vortical semi-dilute dusty flows. These configurations are best suited to tease out the fundamental mechanisms and help build a physical understanding and intuition for how semi-dilute inertial particles modulate flow structures in more complex flows.}

Recently, we have shown that the interaction between disperse inertial particles and a single vortex alters the dynamics from what is commonly understood from particle-free vortex dynamics \citep{shuaiAcceleratedDecayLambOseen2022,shuaiInstabilityDustyVortex2022}. For example, a two-way coupled dusty Lamb-Oseen vortex decays significantly faster than a particle-free one \citep{shuaiAcceleratedDecayLambOseen2022}. This enhanced decay is due to the ejection of inertial particles from the vortical core. While the particles cluster into a ring surrounding the vortex, their feedback force on the fluid leads to faster decay of the flow structure. Perhaps an even more striking effect is the fact that two-way coupled inertial particles dispersed in the core of a two-dimensional vortex trigger an instability \citep{shuaiAcceleratedDecayLambOseen2022}. This is in contrast to the remarkable stability of particle-free vortices to two-dimensional perturbations \citep{fungStabilitySwirlingFlows1975}. We have shown in \citep{shuaiAcceleratedDecayLambOseen2022} that the ejection of the particles from the vortex core activates a centrifugal Rayleigh-Taylor instability that persists even for non-inertial particles.

In light of these previous findings, it is expected that the merger dynamics in two-way coupled dusty flows will be considerably different from those noted in earlier studies of particle-free flows \citep{cerretelliPhysicalMechanismVortex2003a,melanderSymmetricVortexMerger1988,meunierMergingCriterionTwodimensional2002a}. In the present study, we revisit the problem of co-rotating vortices with equal strength, augmented with mono-disperse inertial particles. We use Eulerian-Lagrangian simulations to show new merger mechanisms that depend on particle inertia. A surprising outcome is that inertial particles may even temporarily push apart the two vortices.

{\color{anubhab}
This paper is organized as follows. Section \S \ref{sec:governing_equation} presents the governing equations applied in Eulerian-Lagrangian method to simulate the two-way coupled particle-laden vortices. The vortex merger process in particle-free flow is shown in \S \ref{sec:particle_free}. In \S \ref{sec:particle_laden}, we present the physical mechanism for merger of dusty vortices, the simulation configuration and the numerical results in particle-laden flow. Finally, we provide concluding remarks in \S \ref{conclusion}.
}

\section{Governing equations}\label{sec:governing_equation}
{\color{houssem}
We describe the dynamics of the gas-solid flow using the Eulerian-Lagrangian formalism previously deployed in \citep{shuaiAcceleratedDecayLambOseen2022} and \citep{shuaiInstabilityDustyVortex2022} for the study of vortical dusty flows. For the sake of brevity, we reproduce here only the highlights of the approach.

In the present formulation, the carrier phase is treated in the Eulerian frame, whereas solid particles are tracked individually. The conservation equations for the carrier phase are obtained by volume-filtering the point-wise Navier-Stokes equations \citep{andersonFluidMechanicalDescription1967}. In the semi-dilute regime, the equations for mass and momentum conservation read 
\begin{eqnarray}
\frac{\partial \phi_f}{\partial t}+\nabla\cdot \left(\phi_f\mathbf{u}_f\right) &=&0, \label{eq:2.1}\\
\rho_f\left( \frac{{\partial \phi_f} \mathbf{u}_f}{\partial t}+\nabla\cdot \left(\phi_f\mathbf{u}_f \mathbf{u}_f\right) \right)&=&-\nabla p+\mu_f\nabla ^2 \mathbf{u}_f+\mathbf{F}+\nabla\cdot\mathbf{R}_{\mu},  \label{eq:2.2}
\end{eqnarray}
where $\mathbf{u}_f$ is the fluid velocity, $p$ is the pressure, $\phi_f=1-\phi_p$ is the local fluid volume fraction, $\phi_p$ is the local particle volume fraction, $\rho_f$ is the fluid density, and $\mu_f$ is the fluid viscosity. The tensor $\mathbf{R}_\mu$ results from filtering the point-wise fluid stress tensor and is responsible for the apparent viscosity of the suspension at large particle concentrations $\phi_p>10^{-2}$. This tensor vanishes in the semi-dilute regime considered in this study since $\phi_p=O(10^{-4})$. For the same reason, effects due to volume displacement by the particles are negligible in the semi-dilute regime as $\phi_f\simeq 1$ in equations (\ref{eq:2.1}) and (\ref{eq:2.2}).

The term $\mathbf{F}$ in (\ref{eq:2.2}) represents the momentum exchange between particles and gas. It is due to the feedback force exerted by the particles on the gas and reads
\begin{equation}
\mathbf{F}=  -\phi_p \nabla\cdot \bm{\tau}|_p +\rho_p\phi_p \frac{(\mathbf{u}_p-\mathbf{u}_{f|p})}{\tau_p},\label{eq:coupling}
\end{equation}
where $\bm{\tau}=-p\bm{I}+\mu (\nabla \mathbf{u}_f+\nabla \mathbf{u}_f^T)/2$ is the filtered fluid stress tensor, $\mathbf{u}_p$ is the Eulerian particle velocity, and $\tau_p=\rho_pd_p^2/(18\mu_f)$ is response time of spherical particles with density $\rho_p$ and diameter $d_p$. The subscript ``$(\cdot)_{|p}$'' indicates a quantity evaluated at the particle locations.  The first term in (\ref{eq:coupling}) is the stress exerted by the undisturbed flow. The second term represents the drag force by the particle on the fluid, which is represented using Stokes drag. The latter dominates the interphase coupling (\ref{eq:coupling}) in a dusty gas since the particle-to-fluid density ratio is very large ($\rho_p/\rho_f\gg 1$). 

A scaling analysis of the drag terms shows that the mass loading $M=\rho_p\phi_{p,0}/\rho_f$, where $\phi_{p,0}$ is the average particle volume fraction, determines the strength of the coupling between the gas and solid phases. Thus, for vanishingly small mass loadings, the merger of co-rotating vortices follows the same dynamics as in a particle-free flow since the solid phase has little effect on the gas phase in the limit $M\ll 1$. Conversely, the merger dynamics are expected to deviate from the established dynamics in particle-free flows when $M=O(1)$.

The particle phase is treated in the Lagrangian frame. The motion of the $i$-th particle is given by \citep{maxeyEquationMotionSmall1983}
\begin{eqnarray}
\frac{d\mathbf{x}_p^i}{dt}(t)&=&{\mathbf{u}_p^i}(t),\\
\frac{d\mathbf{u}_p^i}{dt}(t)&=&\frac{1}{\rho_p}\nabla\cdot \bm{\tau}(\mathbf{x}_p^i,t)+ \frac{\mathbf{u}_f(\mathbf{x}_p^i,t)-\mathbf{u}^i_p(t)}{\tau_p},
\end{eqnarray}
where  $\mathbf{x}_p^i$, and $\mathbf{u}_p^i$, are the Lagrangian position and velocity of the ``$i$''-th particle, respectively. Gravity is ignored in this work to focus on  inertial effects. Other interactions, including particle collisions, are negligible due to the large density ratio and low volume fraction in the semi-dilute regime. In the equations above, the instantaneous particle volume fraction and Eulerian particle velocity field are obtained from Lagrangian quantities using
\begin{eqnarray}
\phi_p(\mathbf{x},t)&=&\sum_{i=1}^N V_p  \mathcal{G}\left(\left|\left|\mathbf{x}-\mathbf{x}_p^i\right|\right|\right), \\
\phi_p \mathbf{u}_{f|p}(\mathbf{x},t)&=&\sum_{i=1}^N \mathbf{u}_f(\mathbf{x}_p^i(t),t) V_p \mathcal{G}\left(\left|\left|\mathbf{x}-\mathbf{x}_p^i\right|\right|\right),\\ 
\phi_p \mathbf{u}_p(\mathbf{x},t)&=&\sum_{i=1}^N \mathbf{u}_p^i(t) V_p \mathcal{G}\left(\left|\left|\mathbf{x}-\mathbf{x}_p^i\right|\right|\right),
\end{eqnarray}
where $V_p=\pi d_p^3/6$ is the particle volume, $\mathcal{G}$ represents a Gaussian filter kernel of size $\delta_f=3\Delta x$, where $\Delta x$ is the grid spacing. Further details on the numerical aspects can be found in \citep{capecelatroEulerLagrangeStrategy2013}.}

\section{Particle-free vortex merger}\label{sec:particle_free}

\begin{figure}
  \center
  \includegraphics[width=\linewidth]{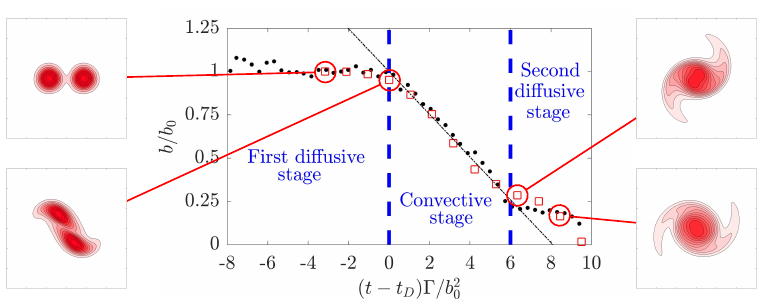}
  \caption{The evolution of vortex separation $b/b_0$ as a function of time in single-phase fluid at $\Rey_\Gamma=530$. Symbols: \FilledSmallCircle, \citet{cerretelliPhysicalMechanismVortex2003a}; \SmallSquare, present simulation. Snapshots represent normalized vorticity.\label{fig:single separation}}
\end{figure}

Before addressing how inertial particles may modulate the merger dynamics, we first describe the different stages of vortex pair merger in a particle-free case that will be used in \S\ref{sec:particle_laden} to assess the effect of introducing inertial particles. {\color{houssem} To this end, we consider a simulation of particle-free vortex merger at circulation Reynolds number $\Rey_{\Gamma}=\rho_f \Gamma/\mu_f=530$, where $\Gamma$ is the vortex circulation, that matches the experiments of \citet{cerretelliPhysicalMechanismVortex2003a}. 

We perform a ``pseudo-2D'' simulation in a periodic domain, subdivided into 4 anti-symmetric quadrants as previously done in \citep{shuaiAcceleratedDecayLambOseen2022} and \citep{shuaiInstabilityDustyVortex2022}. Here, ``pseudo-2D'' refers to the fact that the axial direction is considered periodic and discretized with only one grid point to enable the inclusion of spherical particles at a later point. In each quadrant, we initialize two co-rotating Lamb-Oseen vortices with equal initial radii $a_0$ and separated by $b_0$, such that the ratio $a_0/b_0=0.17$ is initially below the merger threshold $(a/b)_\mathrm{crit}=0.29$. This indicates that the two vortices will not merge immediately, but rather undergo a first diffusive stage before the onset of a convective stage. Each quadrant has a size $30a_0$-by-$30a_0$, which is sufficiently large to avoid interactions between vortices in different quadrants for the length of the simulations considered \citep{shuaiAcceleratedDecayLambOseen2022}. The simulation grid is uniform with a high spatial resolution $a_0/\Delta x \approx 51$  to provide good resolution of the vortex cores. Data collected from all four quadrants is used in the computation of statistics after application of appropriate symmetries.}

According to \citet{cerretelliPhysicalMechanismVortex2003a,melanderSymmetricVortexMerger1988,meunierMergingCriterionTwodimensional2002a}, the merger of two co-rotating vortices follows three stages called the first diffusive stage, the convective stage, and the second diffusive stage. All these stages are reproduced in our simulations as can be seen in figure \ref{fig:single separation}  which show the evolution of the normalized vortex separation $b/b_0$, and axial vorticity $\omega_z$ normalized by the angular velocity $\omega_{\Gamma}=\Gamma/(2\pi a_0^2)$ from one of the four quadrants. Initially, the two vortices rotate around one another, but their separation remains constant as can be seen in figure \ref{fig:single separation}. After a time period $\mathrm{t_D}$, the cores grow sufficiently to reach the merger threshold $(a/b)_\mathrm{crit}=0.29$, at which point, the convective merger is initiated. During this stage, the separation decreases linearly. The vortices are deformed significantly as shown at $(t-t_D)\Gamma/ b_0^2=0$ and 6 in figure \ref{fig:single separation}. The second diffusive stage occurs between  $6\lesssim (t-t_D)\Gamma/ b_0^2\lesssim 9$, after which the vortex pair can be considered fully merged.

\section{Particle-laden vortex merger}\label{sec:particle_laden}

\subsection{The role of disperse phase on vortex merger}
{\color{anubhab}

The merger of vortices in the presence of a particulate phase involves two additional processes -- the dispersion of inertial particles by the background flow and the feedback force from the disperse phase. The former depends on the spatial distribution of strain-dominated regions over rotation-dominated regions, while the latter depends on the local particle concentration and the slip velocity between the two phases. These processes are coupled, and one requires a full numerical simulation to understand completely how a dusty vortex merger behaves differently from its particle-free counterpart. This will be elaborated on in the next section. Here, we will attempt to identify some expected behaviour in dusty vortex mergers based on dispersion and feedback force in an idealized scenario of two liked-signed point vortices of equal circulation. \\

\begin{figure}
\begin{minipage}{0.5\textwidth}
\centering
 \hspace*{-1.8cm}\includegraphics[angle=270,width=3.8in]{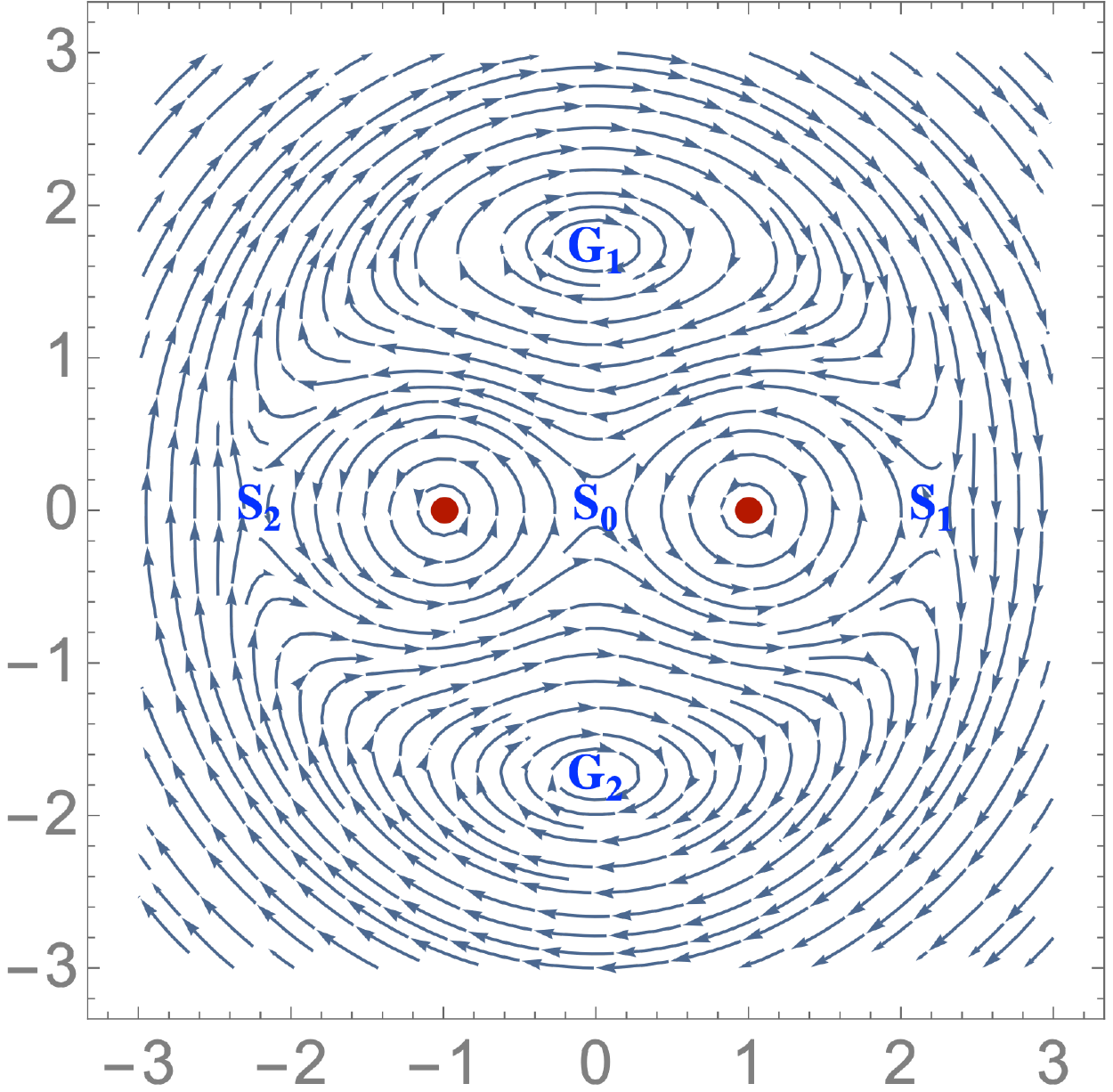}
\caption*{(a) $\mathbf{u}$}
\end{minipage}%
\begin{minipage}{0.5\textwidth}
\centering
 \hspace*{-1.5cm} \includegraphics[angle=270,width=3.8in]{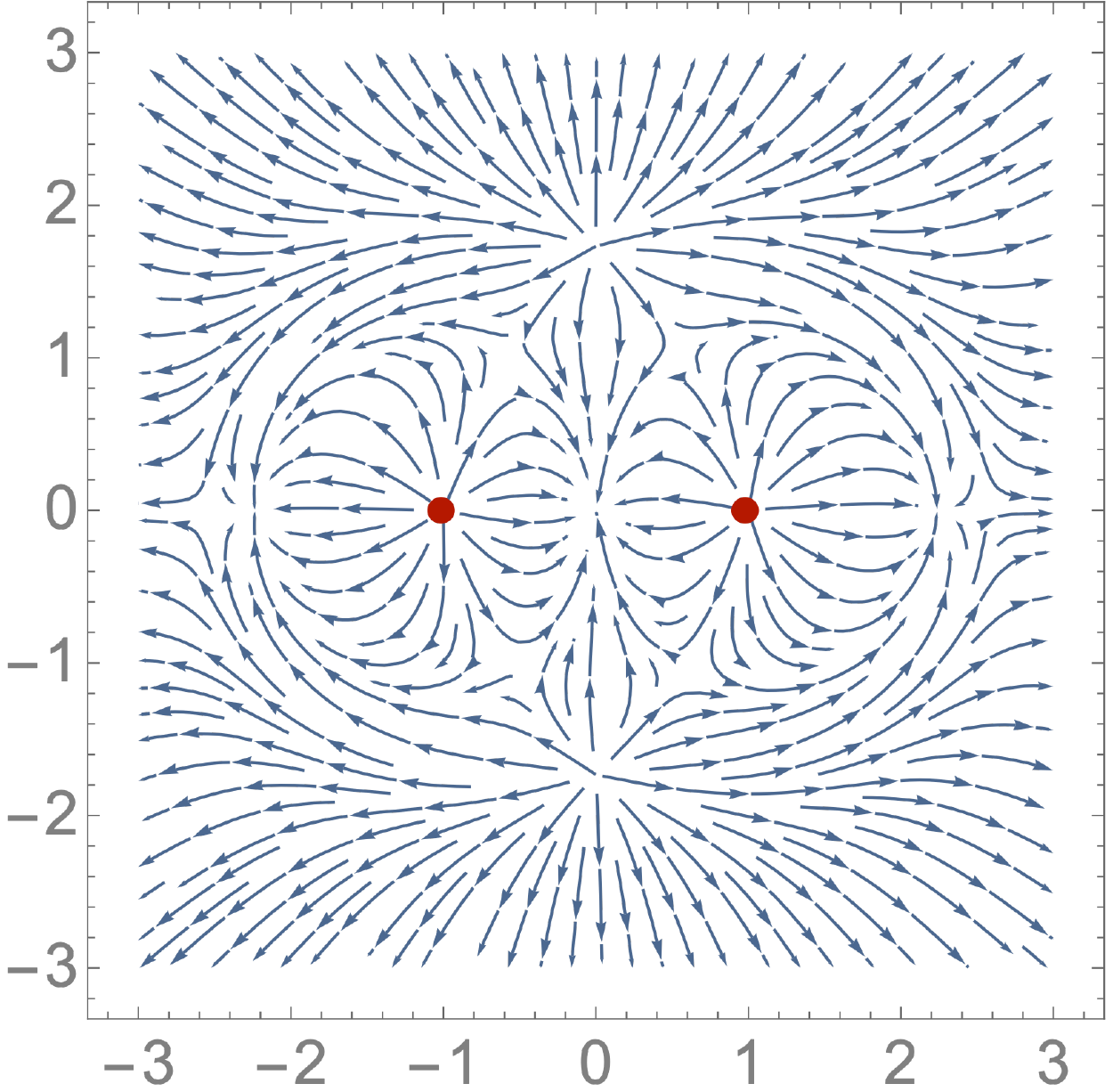}
 \caption*{(b) $-D\mathbf{u}/Dt$}
\end{minipage}
  \caption{Velocity field and negative of the acceleration field ($\propto$ feedback force) for a pair of identical point vortices in co-rotating frame. The point vortices are located at $(\pm 1,0)$ and are marked as red circles. In the velocity field (a) the stagnation points are at $(0,0)$ 
 and $(\pm\sqrt{5},0)$ and are labelled as $\mathbf{S_{0}}$, $\mathbf{S_{1}}$ and $\mathbf{S_{2}}$ respectively. The elliptic fixed points are at $(0,\pm\sqrt{3})$ and are labelled as $\mathbf{G_{1}}$ and $\mathbf{G_{2}}$ respectively.\label{fig:point_vortex}}
\end{figure}
The streamline pattern for a pair of point vortices in the co-rotating frame has three stagnation points ($\mathbf{S_{0}}$, $\mathbf{S_{1}}$ and $\mathbf{S_{2}}$) and two elliptic fixed points ($\mathbf{G_{1}}$ and $\mathbf{G_{2}}$, the counterparts of the ``ghost vortices'' in the finite-sized vortex merger processes) as shown in figure \ref{fig:point_vortex}(a). Heavy inertial particles are expected to centrifuge to infinity in open vortical flows. However, in the flow field for a vortex pair, heavy inertial particles can get trapped in the vicinity of the vortices \citep{angilella2010dust}! The location of fixed point for inertial particles are different from tracers as the ghost vortices, elliptic fixed points for tracers, now turn into stable spirals and migrate towards the saddle points \citep{ravichandran2014attracting}. At a critical Stokes number, the two fixed points merge and disappear. Beyond the critical Stokes number, trapping is no longer possible, and all particles now fly off to infinity. The rate at which particles are centrifuged from the vicinity of the vortices also depends on the Stokes number; lower Stokes would lead to a slower depletion of particles from the neighbourhood of the vortices. \\
\begin{figure}
\begin{minipage}{\textwidth}
\centering
 \includegraphics[width=4in]{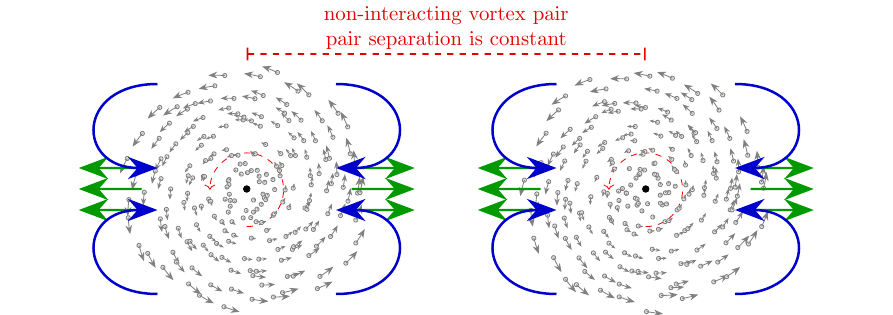}
 \caption*{(a) Low inertia merger}
\end{minipage}\\\\
\begin{minipage}{\textwidth}
\centering
  \includegraphics[width=4in]{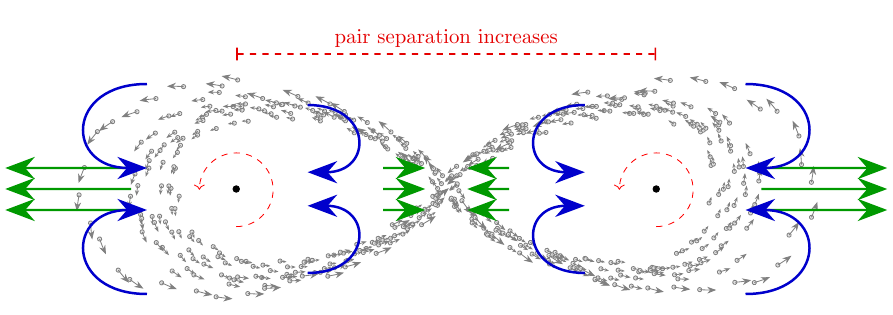}
\caption*{(b) High inertia merger - Initial stage}
\end{minipage}\\\\
\begin{minipage}{\textwidth}
\centering
 \includegraphics[width=4in]{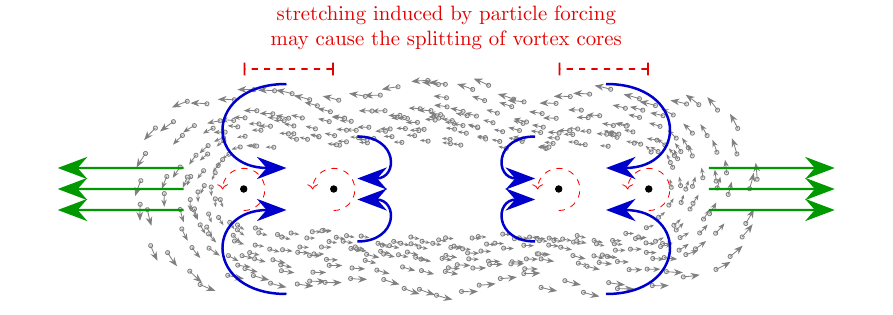}
\caption*{(c) High inertia merger - Later stage}\end{minipage}%
  \caption{Mechanism of vortex merger process in presence of a disperse phase. Green arrows indicate the particle feedback force and the blue arrows indicate the antisymmetric component of the vorticity.\label{fig:schematic}}
\end{figure}
The feedback force of the particles on the fluid depends on the local particle concentration and slip velocity (see equation (\ref{eq:coupling})). In the limit of small particle inertia, the slip velocity can be written as $(\mathbf{u}_p-\mathbf{u}_{f|p})\approx -\tau_pD\mathbf{u}_{f|p}/Dt$, proportional to the local fluid acceleration. The negative of fluid acceleration is plotted in \ref{fig:point_vortex}(b), revealing that each of the vortices experiences a stretching force along the line joining the centers. The forces are in balance when the particle inhomogeneity is minimal outside each vortices. This would be a scenario for cases with low particle inertia, where the merger dynamics would resemble that of particle-free cases. As we will show later in section \ref{low}, the dynamics can be viewed as that of an effective fluid with modified density. The merger dynamics will resemble that shown in the schematic in figure \ref{fig:schematic}(a) -- the particle feedback forces are in balance, and the outer antisymmetric vorticity components are responsible for bringing the vortices together. When particle inertia is higher, the centroid of the two vortex system ($\mathbf{S_{0}}$ in figure \ref{fig:point_vortex}(a)) quickly gets devoid of any particle. The particle depletion near the origin reduces the inward feedback force compared to the outward feedback force, as shown in figure \ref{fig:schematic}(b). Thus, the vortex cores experience an increase in separation due to this imbalance in the feedback force. Once the particles have centrifuged sufficiently far from the vortices, the vortices reverse direction and start approaching each other due to the induced velocity from the antisymmetric vorticity component. For high inertia, the imbalance in the inward and outward feedback forces on each vortex core can also have dramatic effects where they significantly stretch an individual vortex to rip it into two cores as highlighted in figure \ref{fig:schematic}(c). 

Next, we will explore the role of the feedback force and antisymmetric component of vorticity quantitatively using numerical simulations. }

\subsection{Simulation configuration and vortex center identification}
%We now reconsider the merger of the vortex pair in \S\ref{sec:particle_free} under dusty flow conditions, i.e., now, we consider the effects of disperse particles on the merger dynamics.
%To that end, we conduct pseudo-2D Eulerian-Lagrangian simulations with the same methodology previously deployed in \citep{shuaiAcceleratedDecayLambOseen2022} and \citep{shuaiInstabilityDustyVortex2022}. 
Except for the introduction of randomly placed mono-disperse particles, the simulation configuration remains the same as described in \S\ref{sec:particle_free}. The particles have diameter $d_p$, density $\rho_p$, and are initialized with velocities that match the fluid velocity at their locations.

We consider seven cases where the particle inertia and mass loading are varied. Table \ref{tab:table} lists a summary of the non-dimensional parameters in each case. Case A is the reference particle-free case.  Cases B and C correspond to the limit of very low particle inertia, characterized by a circulation Stokes number $\Sto=\tau_p/\tau_f=0.01$. {\color{houssem} Here, $\tau_f=2\pi a_0^2/\Gamma$ is the characteristic fluid timescale associated with an isolated vortex of radius $a_0$ and circulation $\Gamma$}. In these two cases, the mass loading is $\mathrm{M}=1.0$ and 0.5, respectively, obtained by varying the average particle volume fraction $\phi_{p,0}$. In cases D-G, we vary the Stokes number $\Sto$ by changing the particle diameter as shown in Table \ref{tab:table}. For all cases, the initial separation ratio is $a_0/b_0=0.17$ and the density ratio $\rho_p/\rho_f$ is fixed at 2167. \textcolor{houssem}{In order to compute ensemble averages, we repeat the simulations several times, each time with a different realization of the initial random distribution of the particles. The total number of realizations $N_r$ for each case is also shown in table \ref{tab:table}. We chose $N_r$ such that $N_r a/d_p\sim 8,000$, i.e., there are about 16,000 particles within the core of a vortex across all realizations. This ensures that ensemble-averaged quantities are not significantly impacted by discrete fluctuations due to increasing particle size, particularly in cases D-G where $a/d_p<450$.}

\begin{table}
  \begin{tabularx}{\linewidth}{XXXXXX}
  Case  & $\Sto$ &  $M$     & $\phi_{p,0}$ &  $a/d_p$ & \textcolor{houssem}{Number of realizations $N_r$} \\[1ex]
  A     & --          & --         & -- 				 &  --     & \textcolor{houssem}{--}       \\
  B     & 0.01        & 1          & $4.6\times 10^{-4}$ & 1008    & \textcolor{houssem}{8}           \\
  C     & 0.01        & 0.5        & $2.3\times 10^{-4}$ & 1008    & \textcolor{houssem}{8}         \\
  D     & 0.05        & 1          & $4.6\times 10^{-4}$ & 450     & \textcolor{houssem}{8}        \\
  E     & 0.1         & 1          & $4.6\times 10^{-4}$ & 318     & \textcolor{houssem}{8}      \\
  F     & 0.2         & 1          & $4.6\times 10^{-4}$ & 225     & \textcolor{houssem}{32}      \\
  G     & 0.3         & 1          & $4.6\times 10^{-4}$ & 184     & \textcolor{houssem}{32}
  \end{tabularx}
  \caption{Non-dimensional parameters considered in simulations. For all cases, the initial separation ratio is $a_0/b_0=0.17$ and Reynolds number $\Rey_\Gamma=\Gamma/\nu=530$ . }\label{tab:table}
\end{table}

\begin{figure}
 \center
  \includegraphics[width=\linewidth]{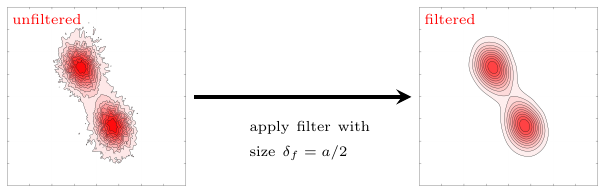}
  \caption{To detect vortex centers, the vorticity field is filtered to remove fluctuations induced by the Lagrangian forcing.\label{fig:filter}}
\end{figure}

The vortex separation is a significant variable to investigate the vortex merger process. Unlike in particle-free merger where smooth vorticity fields allow easy detection of the centers, the feedback force from disperse particles causes large vorticity fluctuations that make the detection of vortex centers harder. The left picture in figure \ref{fig:filter} shows an example of the large vorticity fluctuations obtained in a particle-laden case. Due to this, we have found it necessary to filter the data from Euler-Lagrangian simulations to reliably detect the vortex centers.

To identify the vortex center, we first compute the filtered vorticity $\overline{\omega}_z(\mathbf{x})$ from the instantaneous vorticity field $\omega_z(\mathbf{x})$ by convolving with a triangle filter kernel $g$ with support  $\delta_f$
{\color{houssem}, as follows,
\begin{equation}
  \overline{\omega}_z(x,y,t)=\iint \omega_z(x',y')g(x-x',y-y')dx'dy'.
\end{equation}}
To retain the vortex feature after filtering, $\delta_f$ is set to be half of the initial vortex core radius, that is $a_0/2$. The right picture in figure \ref{fig:filter} shows the result of applying this filtering procedure to the field in the left side of figure \ref{fig:filter}. {\color{houssem} After filtering, we use the gradient descent method to find the local extremums of the filtered vorticity field, thus obtaining the locations of the two vortex centers. The distance between these two centers gives the instantaneous vortex separation $b^{(i)}(t)$ for a realization ``$i$''. Lastly, we ensemble-average results across all realizations to obtain the vortex separation in a dusty gas flow,  $b(t)=(1/N_r)\sum_{i=1}^{N_r}b^{(i)}(t)$}

\subsection{Weakly inertial particles}\label{low}

\begin{figure}
  \center
  \includegraphics[width=5.3in]{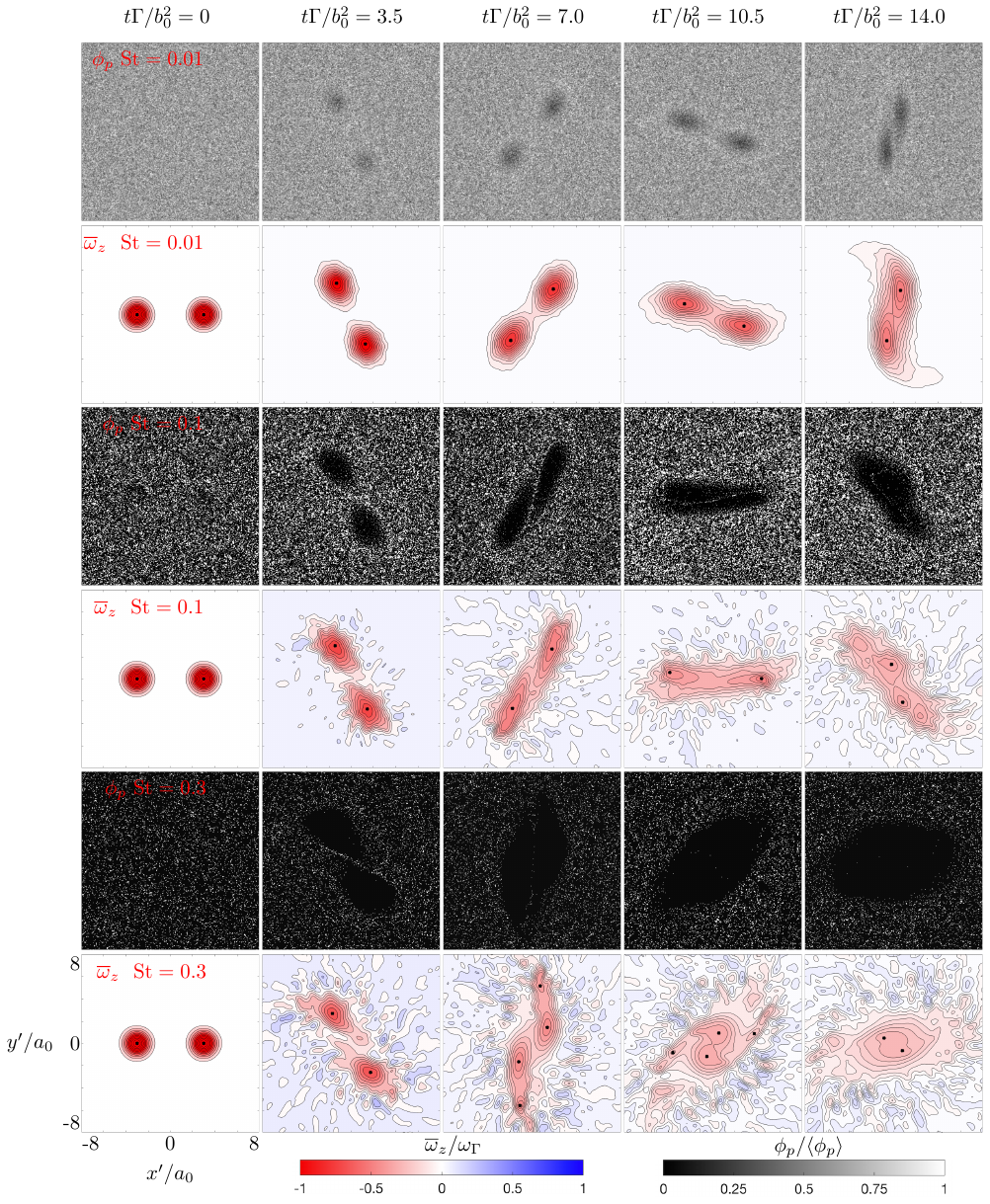}
  \caption{Successive snapshots of normalized particle volume fraction $\phi_p$ and filtered vorticity $\overline{\omega}_z$ iso-contours for cases B, E, and G. See supplementary materials for animations.\label{fig:contour}}
\end{figure}

\begin{figure}
  \centering
  \begin{subfigure}{0.49\linewidth}\centering
    \includegraphics[width=2.4in]{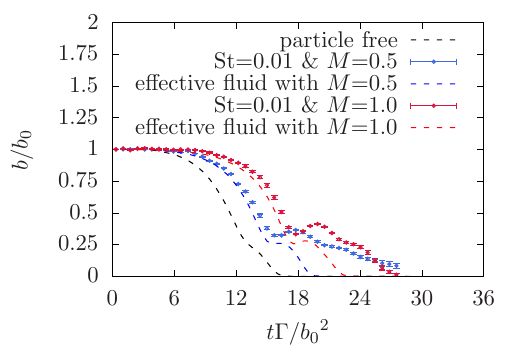}
    \caption{\label{fig:evolution_a}}
  \end{subfigure}
  \begin{subfigure}{0.49\linewidth}\centering
    \includegraphics[width=2.4in]{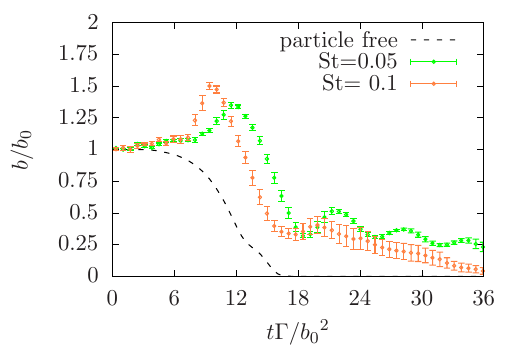}
    \caption{\label{fig:evolution_b}}
  \end{subfigure}\\
  \begin{subfigure}{0.49\linewidth}\centering
    \includegraphics[width=2.4in]{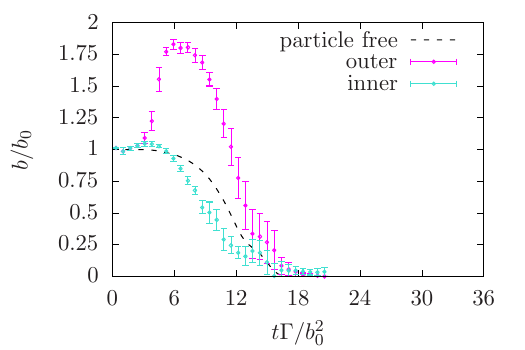}
    \caption{\label{fig:evolution_c}}
  \end{subfigure}
  \begin{subfigure}{0.49\linewidth}\centering
    \includegraphics[width=2.4in]{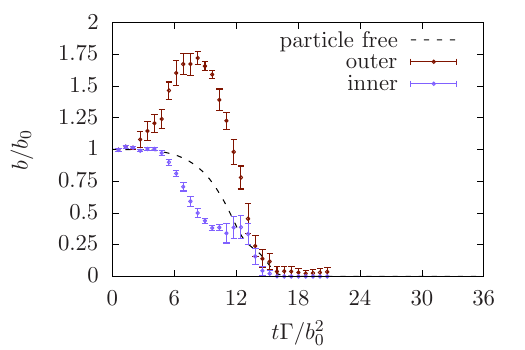}
    \caption{\label{fig:evolution_d}}
  \end{subfigure}
  \caption{Evolution of the normalized separation $b/b_0$ for (a) the weakly inertia cases B-C, (b) the moderately inertial cases D and E, and the highly inertial cases (c) F, and (d) G.
  \label{fig:evolution}} 
\end{figure}
  
Figure \ref{fig:contour} shows the isocontours of normalized particle volume fraction and normalized filtered vorticity from $t\Gamma/b_0^2=0$ to 14 for $\Sto=0.01$, 0.1 and 0.3 with mass loading $M=1$. Videos of these cases are provided in supplementary materials. From the snapshots in figure \ref{fig:contour}, it is clear that semi-dilute inertial particles alter the merger dynamics even at very small Stokes numbers. For the case at $\Sto=0.01$, shown on the first and second rows of figure \ref{fig:contour}, the dynamics of the vorticity field remain qualitatively similar to those of the particle-free vortex pair, however, the merger takes significantly longer and the vortices are further stretched than in the single-phase case. During this process, the particles are gradually ejected from the two vortex cores under the effect of preferential concentration, \textcolor{houssem}{similar to the dynamics observed for an isolated dusty vortex in \citep{shuaiAcceleratedDecayLambOseen2022}}. By $t\Gamma/b_0^2=7$, this results in the formation of two distinct void fraction bubbles.  These structures get progressively larger and more stretched, as can be seen at $t\Gamma/b_0^2=10.5$ and 14.  As the vortices approach one another, a line of particles can be seen dividing the two vortices at $t\Gamma/b_0^2=14$. This line of particles forms because the region between the two vortices is dominated by straining, which draws in particles originally located outside the vortices, and those that have been ejected from the cores.

Figure \ref{fig:evolution_a} shows the evolution of the vortex-pair separation for the particle free case A, and cases B $(\Sto=0.01, M=1.0)$ and C $(\Sto=0.01, M=0.5)$. While the particle-free vortices merge around $t\Gamma/b_0^2\approx 16$, merger occurs much later, around $t\Gamma/b_0^2\approx 28$, with mass loadings $\mathrm{M}=1.0$.

The observation in figures \ref{fig:contour} and \ref{fig:evolution_a} suggest that, at first approximation, the merger of dusty vortex pair at $\Sto\ll 1$ is similar to the merger of particle-free vortices in an effective fluid with density $\rho_\mathrm{eff}=(1+M)\rho_f$. To justify this reasoning, consider a Two-Fluid model of the particle phase where the particle velocity field is expanded in the limit of small inertia as done in \citep{kasbaouiPreferentialConcentrationDriven2015,shuaiInstabilityDustyVortex2022}
\begin{eqnarray}
\frac{\partial \phi}{\partial t} + \bm{u}_p\cdot\nabla \phi &=& -\phi\nabla\cdot\bm{u}_p, \\
\bm{u}_p&=&\bm{u}_f - \tau_p\left(\frac{\partial \bm{u}_f}{\partial t}+\bm{u}_f\cdot\nabla\bm{u}_f\right)+O(\tau_p^2), \label{eq:TF2}
\end{eqnarray}
Combining these equations with the fluid conservation equations,
\begin{eqnarray}
  \nabla\cdot \bm{u}_f=0,\\
  \rho_f \left(\frac{\partial \bm{u}_f}{\partial t}+\bm{u}_f\cdot\nabla\bm{u}_f\right)&=&-\nabla p +\mu \nabla^2\bm{u}_f + \frac{\rho_p \phi}{\tau_p}(\bm{u}_p-\bm{u}_f),\label{eq:TF1}
\end{eqnarray}
yields the following mixture equations
\begin{eqnarray}
  \frac{\partial \rho_\mathrm{eff}}{\partial t} + \bm{u}_f\cdot\nabla \rho_\mathrm{eff}&=& \tau_p \left\{\left(\rho_\mathrm{eff}-\rho_f\right) \nabla\bm{u}_f:\nabla\bm{u}_f+\frac{\nabla\rho_\mathrm{eff}}{\rho_\mathrm{eff}}\cdot\left(-\nabla p +\mu \nabla^2\bm{u}_f\right)\right\}\label{eq:mixture1},\\
  \rho_\mathrm{eff}\left(\frac{\partial \bm{u}_f}{\partial t}+\bm{u}_f\cdot\nabla\bm{u}_f\right)&=&-\nabla p +\mu \nabla^2\bm{u}_f,\label{eq:mixture2}
\end{eqnarray}
where $\rho_\mathrm{eff}=\left(1+M\frac{\phi_p}{\phi_{p,0}}\right)\rho_f$ is the local effective density, the first term on the right-hand side of (\ref{eq:mixture1}) represents preferential concentration and the second term is due to the slip between the two phases. Thus, in the limit of negligible inertia, i.e., $\tau_p\rightarrow 0$, or equivalently, $\Sto\rightarrow 0$, the inertial effects due to preferential concentration and slip vanish making equations (\ref{eq:mixture1}) and (\ref{eq:mixture2}) identical to those of a single-phase fluid with effective density $\rho_\mathrm{eff}$.

To verify this hypothesis, we conducted additional simulations of single-phase merger where the fluid density equals $\rho_\mathrm{eff}=(1+M)\rho_f$ for $M=0.5$ and $M=1.0$. Comparison of the vortex-pair separation from these simulations with the separation measured in the particle-laden cases B and C (see figure \ref{fig:evolution_a}) shows excellent agreement during most of the merger. Deviations that can be seen for $t\Gamma/b_0^2\gtrsim 18$ are likely due to inertial effects which, as suggested by the growth of the void bubbles, become significant as time progresses, despite the low Stokes number $\Sto = 0.01$.

{\color{houssem}
Although the effective fluid analogy captures well the merger dynamics in the limit $\Sto\ll1$, it is worth further investigating the precise mechanisms activated by the particles causing the slowing down of the merger and which may play a larger role with increasing $\Sto$. For this reason, we investigate the dynamics of the ensemble-averaged axial vorticity $\langle\overline{\omega}_z\rangle $ and the particle feedback  $\langle\overline{\bm{F}}\rangle$ as the interplay between these two fields controls the merger. Following \citet{cerretelliPhysicalMechanismVortex2003a}, we investigate these quantities in a rotated reference frame $(x',y')$ such that the $x'$-direction connects the two vortex centers. The $y'$-direction is orthogonal to the latter and represents a plane of symmetry before the vortex cores start deforming. Further, we decompose these quantities into symmetric and antisymmetric parts. For the axial vorticity, this decomposition reads
\begin{eqnarray}
    \langle\overline{\omega}_z\rangle (x',y')&=& \frac{1}{2} \left[\langle\overline{\omega}_z\rangle (x',y') + \langle\overline{\omega}_z\rangle (-x',y')\right] + \frac{1}{2} \left[\langle\overline{\omega}_z\rangle (x',y') - \langle\overline{\omega}_z\rangle (-x',y')\right]\\
    &=&\langle\overline{\omega}_z^S\rangle+\langle\overline{\omega}_z^A\rangle,\label{eq:anti}
\end{eqnarray}
where $\langle\overline{\omega}_z^S\rangle$ and $\langle\overline{\omega}_z^A\rangle$ denote the symmetric and antisymmetric vorticity, respectively. 
For illustration, figure \ref{fig:decompose} shows the filtered  ensemble-averaged vorticity field $\langle\overline{\omega}_z\rangle$ in the laboratory reference frame, in the rotated reference frame, its symmetric part $\langle\overline{\omega}_z^S\rangle$, and antisymmetric part $\langle\overline{\omega}_z^A\rangle$ at an arbitrary time during the merger. 

\begin{figure}
  \centering
  \includegraphics[width=5.2in]{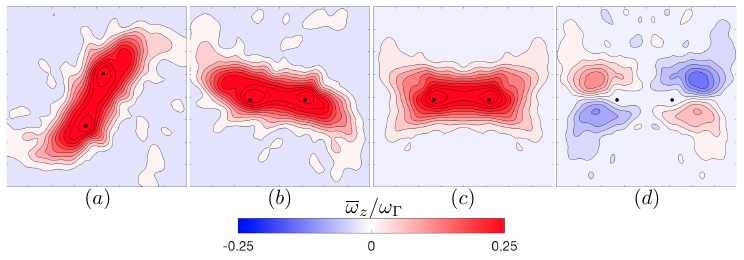}
  \caption{Iso-contours of ensemble-averaged and filtered (a) vorticity $\langle\overline{\omega}_z\rangle(x,y)$ in the laboratory reference frame, (b) vorticity $\langle\overline{\omega}_z\rangle(x',y')$ in the rotated reference frame. (c) symmetric vorticity $\langle\overline{\omega}^S_z\rangle(x',y')$, and (d) antisymmetric vorticity $\langle\overline{\omega}^A_z\rangle(x',y')$.\label{fig:decompose}}
\end{figure}
As argued by \citet{cerretelliPhysicalMechanismVortex2003a}, it is only the antisymmetric vorticity field $\langle\overline{\omega}_z^A\rangle$ that contributes to the change of separation in particle-free cases. Depending on the symmetries of $\langle\overline{\omega}_z^A\rangle$, the induced velocity field may either pull together or push apart the vortex cores. When inertial particles are dispersed in the flow, the antisymmetric part of the component of the particle feedback force in the direction parallel to the line connecting the two vortex centers $\bm{c}_1$ and $\bm{c}_2$, i.e.,
\begin{equation}
  \langle\overline{F}^A_{||}\rangle=\langle\overline{\bm{F}}^A\rangle\cdot \frac{\bm{c}_2-\bm{c}_1}{|| \bm{c}_2-\bm{c}_1||}
\end{equation}
also affects the pair separation. Note that, the dynamics of the symmetric part $\langle\overline{F}_{||}^S\rangle$ in the parallel direction control the translational drift of the whole vortex-pair but do not impact the separation. The dynamics of the symmetric and antisymmetric parts of particle feedback force in the normal direction, i.e., $\langle\overline{F}_{\perp}^S\rangle$ and $\langle\overline{F}_{\perp}^A\rangle$, influence the normal stretching and rotation-rate of the vortex pair, respectively. Since our primary concern is the rate at which a vortex-pair merges in a dusty flow, we focus on analyzing  $\langle\overline{\omega}_{z}^A\rangle$ and $\langle\overline{F}_{||}^A\rangle$ as these are the only two fields impacting the vortex separation.

\begin{figure}
  \center
  \includegraphics[width=5.3in]{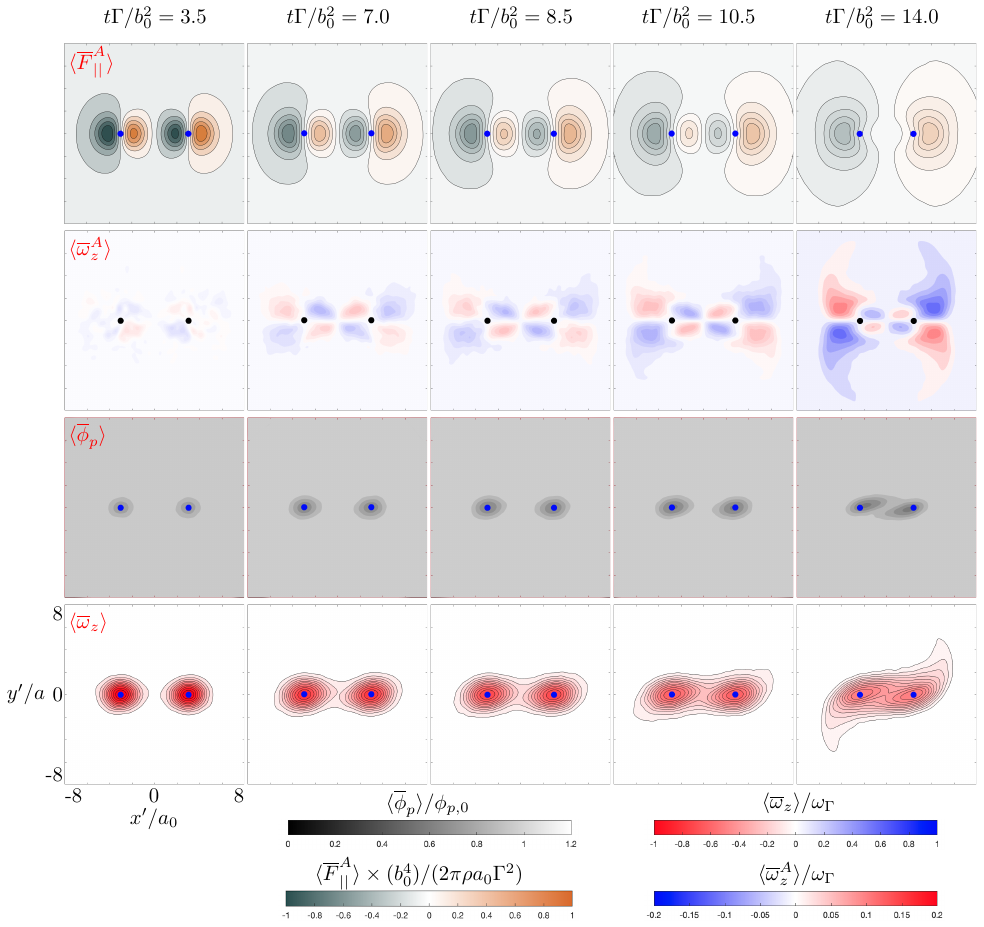}
  \caption{{\color{houssem}Iso-contours of ensemble-averaged and filtered antisymmetric particle-feedback force along the two vortex centers $\langle\overline{F}_{||}^S\rangle$, antisymmetric axial vorticity $\langle\overline{\omega}_{z}^A\rangle$, particle volume fraction $\langle\overline{\phi}_p\rangle$, and total axial vorticity $\langle\overline{\omega}_{z}\rangle$ at representative times during the merger. Data corresponds to case B with  $\Sto=0.01$, $\mathrm{M}=1$, and  $\Rey_\Gamma=530$.} \label{fig:antisymmetric_st001}}
\end{figure}
Figure \ref{fig:antisymmetric_st001} shows the normalized, filtered, and ensemble-averaged particle feedback force in the parallel direction $\langle\overline{F}_{||}^A\rangle$ and antisymmetric vorticity $\langle\overline{\omega}_z^A\rangle$ in the rotated reference frame for case B ($\Sto=0.01$ and $M=1$). To facilitate comparison, the ensemble-averaged particle volume fraction $\langle\overline{\phi}_p\rangle$, fluid vorticity $\langle\overline{\omega}_z\rangle$ in the rotated reference frame are also included in this figure. The antisymmetric vorticity field $\langle\overline{\omega}_z^A\rangle$ displays features that are similar to those observed by \citet{cerretelliPhysicalMechanismVortex2003a} in particle-free vortex merger. Before the convective merger is initiated, at about $\Gamma t/b_0^2\sim 14$, each vortex center is surrounded by two inner and two outer regions where the antisymmetric vorticity is large. The induced velocity by the inner antisymmetric vorticity pushes the two cores apart, while the outer antisymmetric vorticity has the opposite effect of pulling the two cores together. In particle-free merger, there is a balance between these two effects during the first diffusive stage resulting in no change of the separation $b$. With the initiation of the convective merger, the balance between inner and outer antisymmetric vorticity is broken, with the former dominating and causing the separation $b$ to decrease. In the dusty flow case B, figure \ref{fig:antisymmetric_st001} shows that the disperse particles exert a force in the inner region of each core that pulls the two vortices together, and a force on the outer regions that pushes the vortices apart. Later towards $\Gamma t/b_0^2= 14$, the inner attractive force vanishes leaving only the outer force with net repulsive effect. Thus, the particles oppose the attractive pull generated by the antisymmetric vorticity leading to slower merger as observed in figure \ref{fig:evolution_a}.

%Each vortex is subject to two forces with different signs (positive sign means force pointing to the right, negative sign means force pointing to the left). The net particle feedback force exerted on one vortex plays a role in pushing the vortices apart. It is seen that at $t\Gamma/b_0^2=3.5$, 7.0, and 8.5, the net particle force is insignificant as the magnitude difference between the outer force and inner force is small. Meanwhile, the intensity of inner vorticity pairs is close to that of outer pairs.  These little differences for both vorticity intensity and particle forces on two sides result in the stable separation distance between two vortex centers when $t\Gamma/b_0^2 \leq 8.5$ during the first diffusive stage, as shown in figure \ref{fig:evolution_a}. As particles keep migrating from the central area, the particle volume fraction $\overline{\phi}_p$ and the particle forces $\overline{F}_{||,A}$ around the inner region experience a gradual reduction, leading to a larger net particle force exerted on each vortex when $t\Gamma/b_0^2 \ge 10.5$, as shown on the first row and third row in figure \ref{fig:antisymmetric_st001}. Meanwhile, two inner vorticity pairs are smaller than the outer pairs, resulting in a decrease in separation, which is attenuated by the pronounced net particle force pulling the vortex apart. We can conclude that this attenuation causes the merger delay in the case of weakly inertial particles.
}

\subsection{Moderately inertial particles}\label{moderate}

While the merger dynamics of a dusty flow with weakly inertial particles ($\Sto\ll 1$)  are qualitatively similar to those of a particle-free flow, new dynamics emerge with increasing particle inertia. The most notable change noted in our simulations with  $0.05 \leq \Sto \leq 0.1$ is that the eventual merger of the vortex pair starts first by the two vortices pushing apart.

The void bubbles in case E ($\Sto = 0.1$, $M=1.0$), shown in figure \ref{fig:contour}, grow significantly faster than in the low inertia case B,  as the effects of preferential concentration intensify with increasing particle inertia \citep{shuaiAcceleratedDecayLambOseen2022}.  Further, the deformation of vortex cores and the void bubbles starts earlier suggesting that this process is related to particle inertia. Due to the faster depletion of the cores, the particle line separating the two vortices appears earlier, at around $t\Gamma/b_0^2=7$, and becomes thinner as the merger progresses. During this early transient $t\Gamma/b_0^2\lesssim 10.5$, the two cores push apart leading to an increase in separation compared to the initial state. The cores start approaching one another only once the line of particles separating the cores becomes sufficiently thin, which eventually ruptures.

Figure \ref{fig:evolution_b} shows the evolution of the normalized separation $b/b_0$ for cases D and E, alongside the data for the particle-free case A. While $t\Gamma/b_0^2\lesssim 7$, the separation remains approximately constant. During this stage, the two vortices are mostly independent from one another and evolve according to dynamics similar to those reported in \citep{shuaiAcceleratedDecayLambOseen2022}. Unlike in single-phase merger where the growth of the vortex cores is exclusively driven by viscosity, the growth of the cores and void bubbles are interlinked as the feedback force from the particles exiting the cores causes greater spreading of the vorticity field. {\color{houssem} Time $t\Gamma/b_0^2\simeq 7$ marks the start of a new stage, that we call \emph{repulsion stage}, and which ends by $t\Gamma/b_0^2\simeq 12$ in case D ($\Sto =0.05$) and $t\Gamma/b_0^2\simeq 10$ in case  E ($\Sto =0.1$). During this stage, the vortex pair separation increases monotonically, up to a saturation limit $b/b_0\ \sim 1.35$ and 1.5 in cases D and E, respectively. Then, the merger is initiated. This stage resembles the convective stage in particle-free merger, during which the vortex pair separation drops rapidly and lasts until $t\Gamma/b_0^2\simeq 18$ in case D and $t\Gamma/b_0^2\simeq 16$ in case E. At the end of this stage,  the two void regions have merged, resulting in a large particle-free region containing the two vortices. The dynamics from here and onward follow those of single-phase merger as the particles have been ejected from the central region.}

{\color{shuai}

\begin{figure}
  \center
  \includegraphics[width=5.3in]{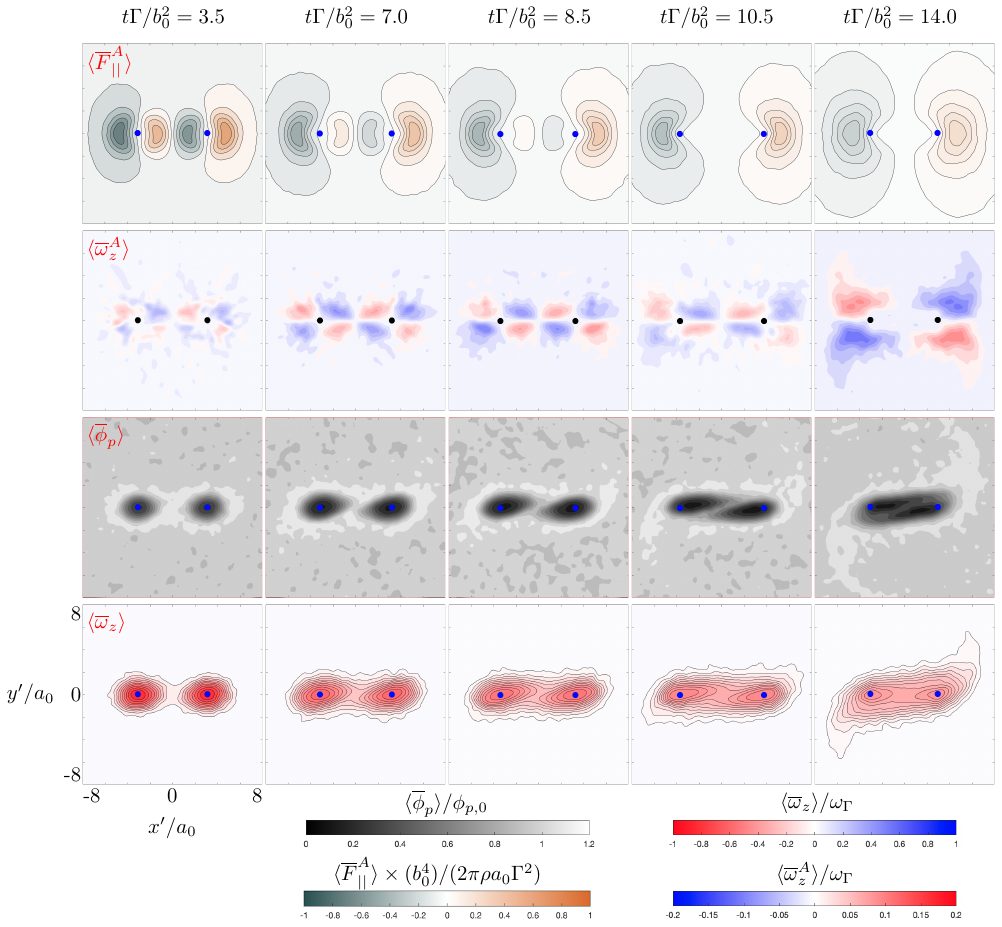}
  \caption{{\color{houssem}Iso-contours of ensemble-averaged and filtered antisymmetric particle-feedback force along the two vortex centers $\langle\overline{F}_{||}^S\rangle$, antisymmetric axial vorticity $\langle\overline{\omega}_{z}^A\rangle$, particle volume fraction $\langle\overline{\phi}_p\rangle$, and total axial vorticity $\langle\overline{\omega}_{z}\rangle$ at representative times during the merger. Data corresponds to case D with  $\Sto=0.05$, $\mathrm{M}=1$, and  $\Rey_\Gamma=530$.} \label{fig:antisymmetric_st005}}
\end{figure}
}

{\color{houssem} 
To elucidate the mechanism driving the repulsion stage, we report in figure \ref{fig:antisymmetric_st005} isocontours of the fields $\langle \overline{\omega}_z^A\rangle$, $\langle\overline{F}_{||}^A\rangle$, $\langle \overline{\phi}_p\rangle$ and $\langle \overline{\omega}_z\rangle$ for case D ($\Sto=0.05$) at representative times in the rotated reference frame. As previously discussed in \S\ref{low}, it is the symmetries of $\langle \overline{\omega}_z^A\rangle$ and $\langle\overline{F}_{||}^A\rangle$, and their interplay, that dictate the evolution of the vortex pair separation. Up until $t\Gamma/b_0^2 \simeq 10.5$, there is a relative balance between the inner and outer antisymmetric vorticity $t\Gamma/b_0^2 = 10.5$. This suggests that the vorticity dynamics do not have a significant effect on the vortex pair separation during this time. In contrast, a gradual imbalance develops between the inner and outer parts of the parallel antisymmetric particle feedback force $\langle\overline{F}_{||}^A\rangle$. This imbalance favors the outer regions which have a net repulsive effect on the vortex pair. It is caused by the drag force exerted by inertial particles ejected away from the vortex pair. The weakening of the particle feedback force in the inner regions results from the growth of the void fraction bubbles and their gradual merger. This imbalance leads to a gradual increase of the vortex pair separation. This effect accelerates significantly at about $t\Gamma/b_0^2 = 10.5$, which  represents the time around which the inner regions become fully depleted from particles and no longer exert any pull on the vortex pair.
Later, the antisymmetric vorticity develops an imbalance between inner and outer regions in turn, which can be seen in the isocontours at $t\Gamma/b_0^2 = 14$. The induced velocity by the breaking of this balance causes the vortex cores to pull together. When this attractive effect of the antisymmetric vorticity overcomes the repulsive effect of the disperse particles,  merger is initiated and the vortex pair separation decreases rapidly.

}

\subsection{Highly inertial particles}\label{high}

\begin{figure}
  \center
  \includegraphics[width=5.3in]{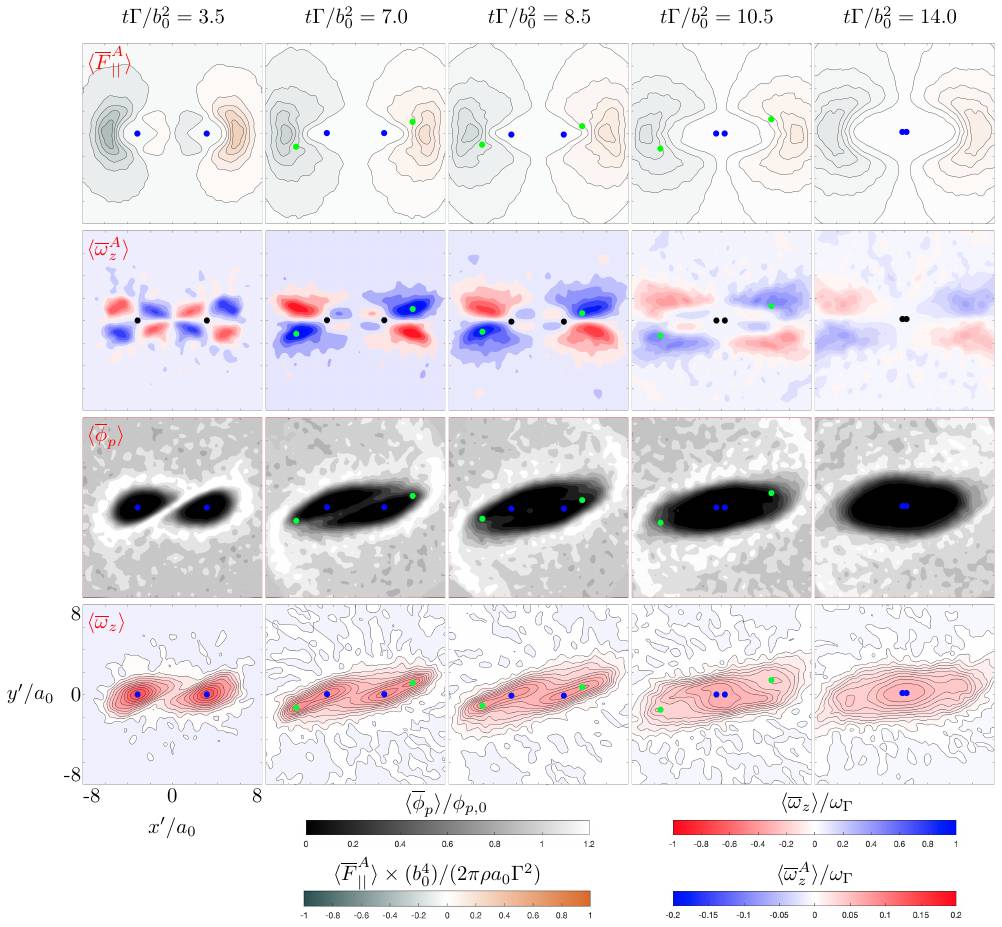}
  \caption{{\color{houssem}Iso-contours of ensemble-averaged and filtered antisymmetric particle-feedback force along the two vortex centers $\langle\overline{F}_{||}^S\rangle$, antisymmetric axial vorticity $\langle\overline{\omega}_{z}^A\rangle$, particle volume fraction $\langle\overline{\phi}_p\rangle$, and total axial vorticity $\langle\overline{\omega}_{z}\rangle$ at representative times during the merger. Data corresponds to case F with  $\Sto=0.2$, $\mathrm{M}=1$, and  $\Rey_\Gamma=530$.} \label{fig:antisymmetric_st02}}
\end{figure}

With increasing Stokes number, the feedback force from the particles increasingly distorts the vortical structures making the merger \textcolor{houssem}{more complex}. This is illustrated in case G ($\Sto=0.3$, $M=1$), in figure \ref{fig:contour}, where the vortices appear highly stretched at a time as early as $t\Gamma/b_0^2 =3.5$. This extreme distortion causes each vortex to split into two smaller vortices, an inner one and an outer one, as can be clearly seen at $t\Gamma/b_0^2 = 7$ \textcolor{houssem}{in the instantaneous fields in figure \ref{fig:contour} and the ensemble-averaged fields in figure \ref{fig:antisymmetric_st02}}. From figure \ref{fig:evolution_c} and \ref{fig:evolution_d}, the inner vortices start merging around $t\Gamma/b_0^2 = 18$ for case F ($\Sto=0.2$) and $t\Gamma/b_0^2 = 14$ for case G ($\Sto=0.3$), with no repulsion stage. Meanwhile, the outer vortices start with a repulsive stage for $3\lesssim t\Gamma/b_0^2 \lesssim 9$, during which the centers push apart to a maximum distance $b/b_0\simeq 1.9$. This stage is followed by a convective stage and a second diffusive stage between $9\lesssim t\Gamma/b_0^2 \simeq 16$. At the end, a single distorted vortex is left, enclosed inside a larger void fraction bubble.

\begin{figure}
  \centering
  \begin{subfigure}{0.45\linewidth}\centering
    \includegraphics[width=2.4in]{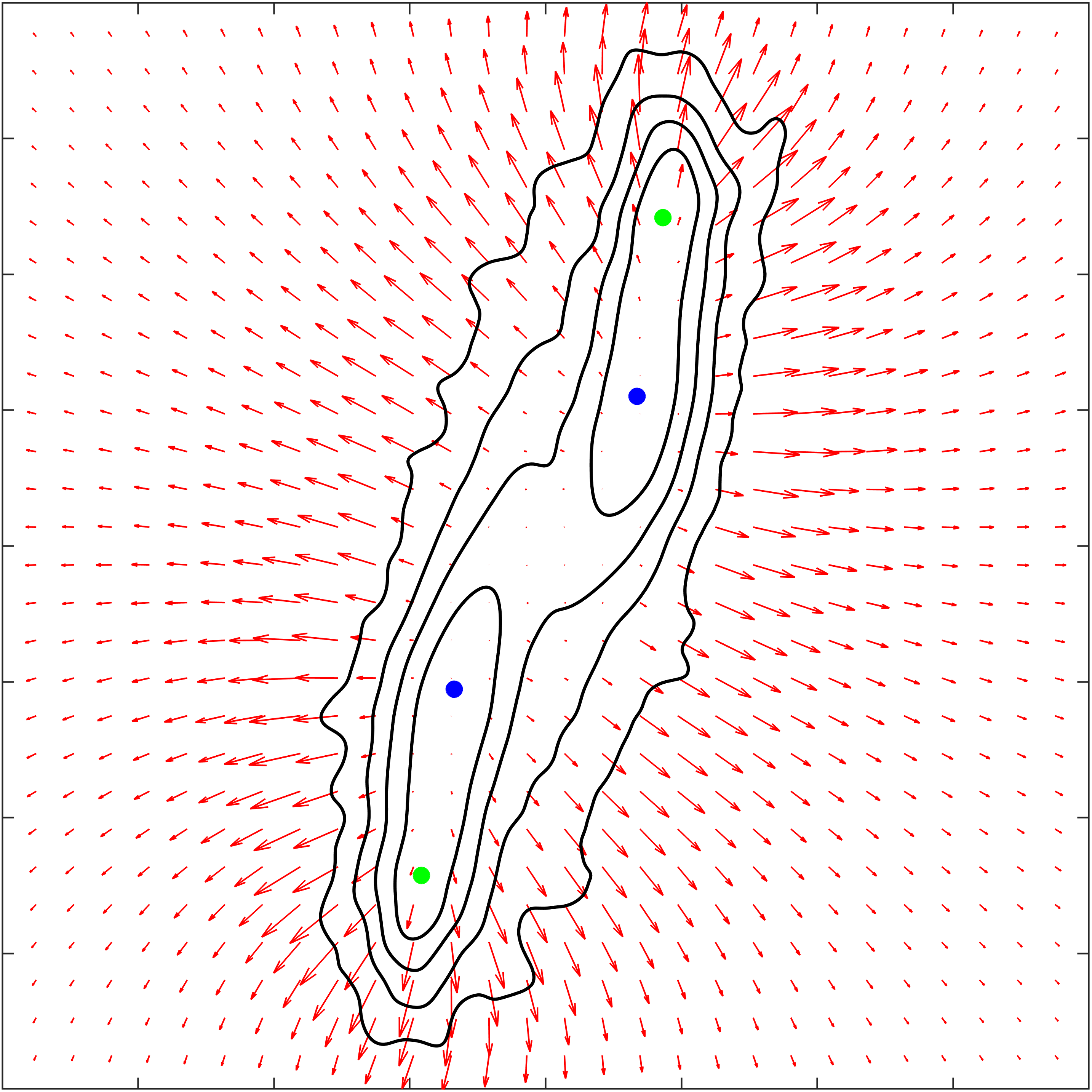}
    \caption{}
  \end{subfigure}
  \begin{subfigure}{0.45\linewidth}\centering
    \includegraphics[width=2.4in]{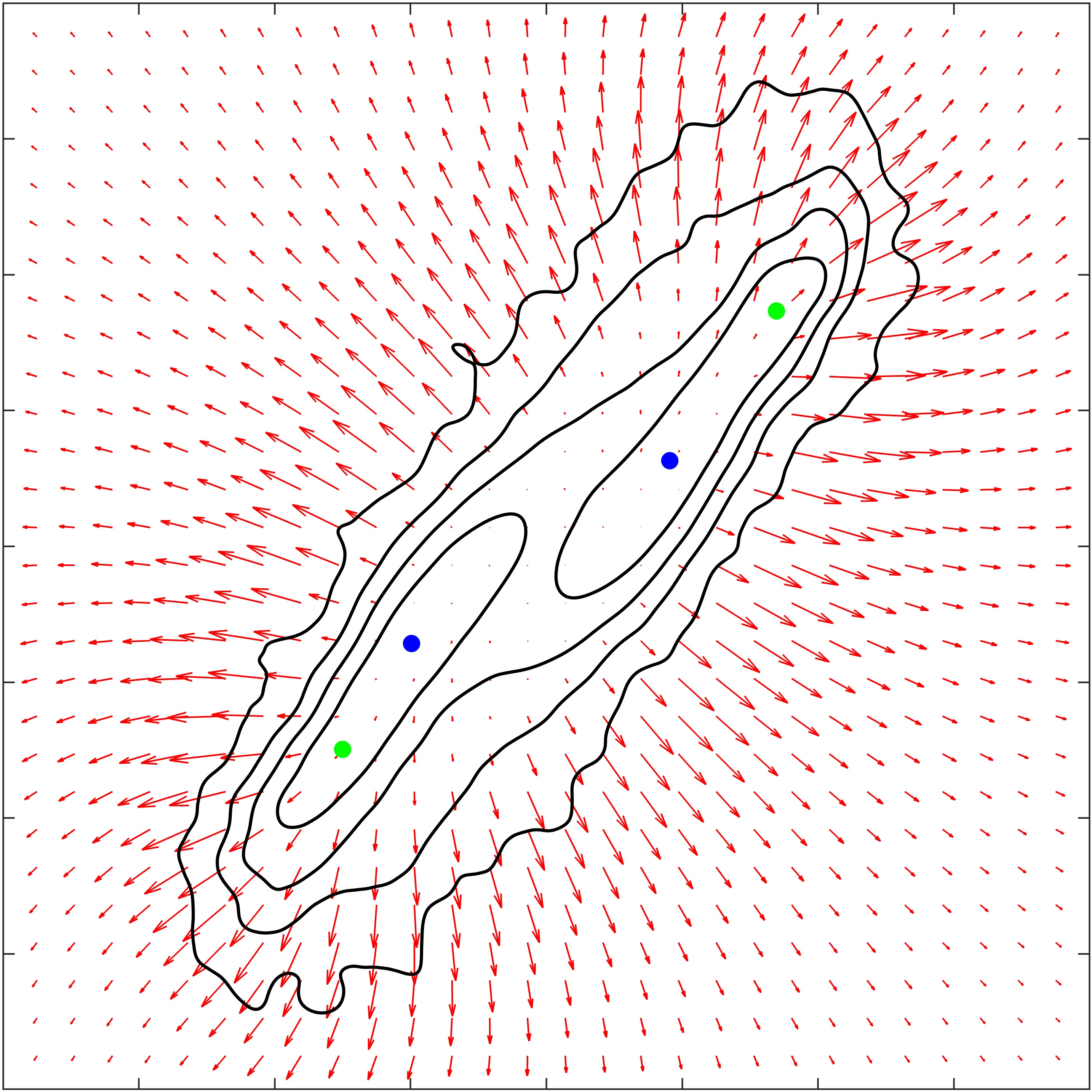}
    \caption{}
  \end{subfigure}\\
  \caption{{\color{houssem}Vector plot of  $\langle\overline{\bm{F}}\rangle$ overlayed on isocontours of axial vorticity $\langle\omega_z\rangle$ from case F ($\Sto=0.2$, $\mathrm{M}=1$) at non-dimensional times (a) $t\Gamma/b_0^2=7$ and (b) $t\Gamma/b_0^2=7$.  The vectors are scaled by magnitude.}\label{fig:vector_fields} }
\end{figure}

{\color{houssem}
The splitting of each vortex results from the extreme distortion caused by the particles. Figure \ref{fig:antisymmetric_st02} shows that the initial vortex pairs stretches under the influence of the disperse particles. By $t\Gamma/b_0^2=3.5$, the vortices assume elliptical shapes with similarly shaped void-fraction bubbles. Due to the fast depletion of the inner region between the two cores, the particle feedback force has a net repulsive effect on the vortex pair. However, since this force is largest at the opposite ends of the vortex pair, it causes significant stretching of the cores and ultimately causes the appearance of two vorticity extrema for each initial vortex core. 

The inner vortices start merging when they become decoupled from the disperse particles. This occurs at around $t\Gamma/b_0^2=7.0$ as the inner region is devoid of particles at this point. Figure \ref{fig:vector_fields} shows vector fields, where vectors are scaled by magnitude, of the ensemble-averaged particle forcing $\langle\overline{\bm{F}}\rangle$ in the laboratory reference frame at times $t\Gamma/b_0^2=7.0$ and $t\Gamma/b_0^2=8.5$. Whereas the outer vortices are dragged outward by the centrifuging particles, the inner vortices are not subject to any forcing due to the absence of particles locally. Meanwhile, the induced velocity by the inner vortices pulls them together, following similar dynamics to those acting during the convective merger of particle-free vortices, as can be seen from the antisymmetric vorticity in figure \ref{fig:antisymmetric_st02} at times $t\Gamma/b_0^2=7.0$ and $t\Gamma/b_0^2=8.5$. Ultimately, this causes the inner vortices to merge before the outer ones.

The merger of the outer vortices starts around $t\Gamma/b_0^2\sim 9$ and concludes around $t\Gamma/b_0^2\sim 18$ and $t\Gamma/b_0^2\sim 16$ for cases F and G, respectively.  The outer vortices merge when the antisymmetric vorticity exerts a greater pull than the repulsion caused by the particles, following dynamics similar to those noted in \S \ref{moderate}.
}

%\begin{figure}
%  \centering
%  \begin{subfigure}{0.45\linewidth}\centering
%    \includegraphics[width=2.4in]{figures/contour_020.png}
%    \caption{\label{fig:force_st02_20_a}}
%  \end{subfigure}
%  \begin{subfigure}{0.45\linewidth}\centering
%    \includegraphics[width=2.4in]{figures/Force_020.png}
%    \caption{\label{fig:force_st02_20_b}}
%  \end{subfigure}\\
%  \caption{{\color{shuai}Iso-contours of vorticity $\overline{\omega}$ and the filtered particle force field for the case of $\Sto=0.2$, at non-dimensional times $t\Gamma/b_0^2=7.0$, $\Rey_\Gamma=530$. $\mathrm{M}=1$} \label{fig:force_st02_20} } 
%\end{figure}
%
%\begin{figure}
%  \centering
%  \begin{subfigure}{0.45\linewidth}\centering
%    \includegraphics[width=2.4in]{figures/contour_024.png}
%    \caption{\label{fig:fig:force_st02_24_a}}
%  \end{subfigure}
%  \begin{subfigure}{0.45\linewidth}\centering
%    \includegraphics[width=2.4in]{figures/Force_024.png}
%    \caption{\label{fig:fig:force_st02_24_b}}
%  \end{subfigure}\\
%  \caption{{\color{shuai}Iso-contours of vorticity $\overline{\omega}$ and the filtered particle force field for the case of $\Sto=0.2$, at non-dimensional times $t\Gamma/b_0^2=8.5$, $\Rey_\Gamma=530$. $\mathrm{M}=1$} \label{fig:force_st02_24} }
%\end{figure}

\section{Conclusion}\label{conclusion}

Eulerian-Lagrangian simulations of the merger of co-rotating vortices laden with inertial particles reveal new mechanics specific to dusty flows. The present simulations were carried out in the semi-dilute regime, specifically for average particle volume fractions $\phi_{p,0}=2.3 - 4.6 \times 10^{-4}$ and mass loading $M=0.5 - 1.0$. Despite the low particle concentration, dusty flows in this regime have strong momentum coupling between the carrier and disperse phase since the mass loading is order unity. To investigate the effect of particle inertia, we varied the Stokes number $\Sto$ in the range 0.01 to 0.3. We found that these particles can be classified into three main categories. Particles that have a Stokes number $\Sto\leq 0.01$ are considered weakly inertial. With such particles, the merger of dusty vortices is delayed compared to the merger of particle-free vortices. However, the merger dynamics are not much different from those of the particle-free case, if one considers the particle-fluid mixture as an effective fluid with density $\rho_\mathrm{eff}=(1+M)\rho_f$. Particles with Stokes number in the range $\sim 0.05$ to $\sim 0.1$ are considered as moderately inertial. In this case, the merger of a dusty vortex pair exhibits an additional stage characterized by a temporary repulsion of the vortex cores, before undergoing successive convective merger and second diffusive stages. Analyzing the antisymmetric vorticity field \textcolor{houssem}{and the antisymmetric particle feedback force in a corotating reference frame}, we find that the vortex separation increase is caused by a repulsive force generated by the ejection of particles from the vortex cores. Once all particles have been ejected from the inner region separating the two cores, {\color{houssem}the attractive effect of the antisymmetric vorticity field dominates, which triggers the merger of the vortex pair. For highly inertial particles ($\Sto\gtrsim 0.2$), the particle feedback force causes each core to stretch to such extent that it rips it into two cores. In this case, the merger of the inner and outer vortices takes place in sequence. The inner vortices initiate the merger as soon as the inner region becomes devoid of particles. The outer vortices push apart temporarily and then initiate the merger once the inertial particles are ejected sufficiently away. In all cases, the final outcome of the merger is a single vortex with a core that is depleted from all particles and a surrounding hallo of high particle concentration.}

\vspace{1ex}

\noindent\textbf{Acknowledgments.} The authors acknowledge support from the US National Science Foundation (award \#2148710, CBET-PMP), and from IIT Madras for the ``Geophysical Flows Lab'' research initiative under the Institute of Eminence framework.

\noindent\textbf{Declaration of Interests.} The authors report no conflict of interest.

\bibliography{references/shuai.bib,references/houssem.bib}
\end{document}